\documentclass[onecolumn,11pt,a4paper,aps]{revtex4}

\newcommand{\eq}[1]{\begin{equation}#1\end{equation}}
\newcommand{\dd}{\mathrm{d}}
\newcommand{\ee}{\mathrm{e}}

\newcommand{\ch}{\mathrm{ch\,}}

\usepackage{amsmath}
\usepackage{graphics}
\usepackage{graphicx}
\usepackage{psfrag}

\begin{document}

\title{On entanglement evolution across defects in
critical chains}
\author{Viktor Eisler$^1$ and Ingo Peschel$^2$}
\affiliation{
$^1$Vienna Center for Quantum Science and Technology,\\
Faculty of Physics, University of Vienna,
Boltzmanngasse 5, A-1090 Wien, Austria\\
$^2$Fachbereich Physik, Freie Universit\"at Berlin,
Arnimallee 14, D-14195 Berlin, Germany\\
}

\begin{abstract}
We consider a local quench where two free-fermion half-chains are coupled via 
a defect. We show that the logarithmic increase of the entanglement entropy is governed
by the same effective central charge which appears in the ground-state properties and 
which is known exactly. For unequal initial filling of the half-chains, we determine 
the linear increase of the entanglement entropy. 
\end{abstract}

\maketitle

\section{Introduction}

The entanglement across defects in critical quantum chains is interesting, because in
free-fermion systems it varies continuously with the defect strength \cite{Peschel05}. For
two half-chains of length $L$, one finds a logarithmic behaviour of the von Neumann
entanglement entropy
\begin{equation}
 S = \frac{c_{\mathrm{eff}}}{6} \ln L
\label{static}
\end{equation}
where $0 \le c_{\mathrm{eff}} \le c$ with $c$ (equal to 1/2 or 1) denoting the central 
charge of the model.
Since $e^S$, which is the effective number of terms in the Schmidt decomposition, 
then follows a power law, one can view $c_{\mathrm{eff}}$ as a variable critical
exponent. It was studied in a number of papers for discrete \cite{Peschel05,Igloi/Szatmari/Lin09,
Eisler/Peschel10,Eisler/Garmon10,Levine/Friedman11} and continuous \cite{CMV11,CMV12a,CMV12b}
systems and an exact analytical formula was obtained \cite{Eisler/Peschel10}, also for the 
R\'enyi entropy \cite{CMV12a,Peschel/Eisler12} and for 
bosons \cite{Sakai/Satoh08,Peschel/Eisler12}. Technically, the variation is caused by 
a gap in the single-particle 
eigenvalue spectrum of the reduced density matrix (RDM) and the relevant physical parameter
is the transmission amplitude of the defect.

A similar situation is found if one considers a local quench where one connects two 
initially separated half-chains by a defect. The entanglement entropy then increases
in time as 
\begin{equation}
 S = \frac{\hat c_{\mathrm{eff}}}{3} \ln t
\label{dynamic}
\end{equation}
as long as $t \ll L$ for finite $L$ and for all times if $L$ is infinite. This was found
either purely numerically \cite{Igloi/Szatmari/Lin09} or by evaluating expressions based 
on the counting statistics numerically \cite{Klich/Levitov09,Song/etal11,Song/etal12}. No explicit 
formulae for $\hat c_{\mathrm{eff}}$ have been given so far. However, it was observed in the 
numerics of the transverse Ising model that $\hat c_{\mathrm{eff}}$ equals $c_{\mathrm{eff}}$, 
as in the the homogeneous case, where the same $c$ appears in (\ref{static}) and (\ref{dynamic}) 
and the formulae follow from conformal considerations \cite{CC09}.

In this note, we look at this problem once again and study the evolution after the quench
for an XX (hopping) model via the time-dependent single-particle RDM eigenvalues.
We show that those for the defect case are related by a simple exact formula to those 
for the homogeneous case, if the defect is conformal (i.e. scale-free). The relation is 
the same as in the static case and thus implies again a gap in the spectrum.
This allows to take over the ground-state results and thereby proves the equality of
$c_{\mathrm{eff}}$  and $\hat c_{\mathrm{eff}}$. Since $c_{\mathrm{eff}}$ is exactly
known, it also provides an analytical formula for $\hat c_{\mathrm{eff}}$. One can check 
that it fully agrees with the expressions and results of the counting statistics.

We also consider the case where the initial filling of the two half-chains
is different. Then a somewhat different relation for the spectra exists where the
effect of backscattering is manifest. The current which sets in after the quench
leads to a steady generation of entanglement between transmitted and reflected
parts of the wavefunction and thus to a linear increase of the entropy with time.
We give analytical expressions for the coefficient both for conformal and for 
non-conformal defects.

\section{Model and method}

We study free fermions hopping on a finite open chain of $2L$ sites.
The time evolution for $t>0$ is governed by the Hamiltonian
\eq{
H' = \frac{1}{2}\sum_{m,n=1}^{2L} H'_{m,n} c_m^\dag c_n
\label{eq:ham}}
where the nonzero matrix elements are
\eq{
H'_{m,m+1}= H'_{m+1,m} = \left\{
\begin{array}{ll}
-1 & \quad m \ne L \\
-\lambda  & \quad m = L
\end{array}\right. , \quad
H'_{L,L} = -H'_{L+1,L+1}=\sqrt{1-\lambda^2}
\label{eq:hmn}}
In the middle of the chain there is a defect in the form of a modified bond supplemented 
with site energies on both sides and characterized by the parameter $\lambda$.
It will be referred to as the conformal defect 
and is used to derive all the exact relations. In addition, we will also consider simple 
bond defects without site energies. The initial system is composed of two disconnected half-chains 
and the initial state will be specified in the  corresponding sections.

In the homogeneous case, $\lambda=1$, the Hamiltonian (\ref{eq:ham})
is diagonalized by a Fourier transform and the eigenvectors and eigenvalues are given by
\eq{
\phi_k(m) = \sqrt{\frac{2}{2L+1}}\sin\frac{\pi k m }{2L+1}, \quad
\Omega_k = -\cos \frac{\pi k}{2L+1}
\label{eq:eigv}
}
where $k = 1,\dots,2L$.
For the conformal defect the solutions are related to those of the homogeneous chain via
\eq{
\phi'_k(m) = \left\{
\begin{array}{ll}
\alpha_k \phi_k(m) & \quad 1 \le m \le L \\
\beta_k \phi_k(m) & \quad L <  m \le 2L
\end{array}\right. , \quad
\Omega'_k = \Omega_k
\label{eq:eigvcd}
}
and thus the eigenvectors are simply rescaled on the left and right hand side of the defect
with the scaling factors
\eq{
\alpha_k^2 = 1 + (-1)^k \sqrt{1-\lambda^2}, \quad
\beta_k^2 = 1 - (-1)^k \sqrt{1-\lambda^2}
\label{eq:ab}
}
Note that, apart from the alternation, $\alpha_k$ and $\beta_k$ are independent of 
$k$ and lead to a constant transmission coefficient $T=\lambda^2$. Therefore
one has here a lattice realization of the special scale-free defects which can be constructed
by gluing together two conformal field theories \cite{Bachas07,Peschel/Eisler12}. 
The bond defect is non-conformal 
and leads to a transmission which depends on the wavelength of the incoming particle.

For the entanglement between left and right halves, one needs the RDM for a half-chain
which has the form $\rho = \ee^{-\mathcal{H}}/Z$ with a free-fermion effective Hamiltonian 
$\mathcal{H}$, see \cite{review09}. In the following, the single-particle eigenvalues 
of $\mathcal{H}$ will be called $2\omega_l(t)$.
They are related via
\eq{
\zeta'_l(t)=\frac{1}{\ee^{2\omega_l(t)}+1}
}
to the eigenvalues of the
correlation matrix $\mathbf{C}'(t)$ which is the full matrix $\mathbf{\bar C}'(t)$ with  
elements $\langle c^\dag_m(t) c_n(t)\rangle$ restricted to $1\le m,n \le L$.
For $\lambda=1$, the quench will be called \emph{homogeneous} and the corresponding quantities 
are denoted by $2\varepsilon_l(t)$, $\zeta_l(t)$ and $\mathbf{C}(t)$, respectively. 

Therefore one has to find the time-dependent correlation matrix which can be written 
in the Heisenberg form
\eq{
\mathbf{\bar C}'(t) = \ee^{i \mathbf{\bar H}'t} \mathbf{\bar C} (0) 
\ee^{-i \mathbf{\bar H}'t}
\label{eq:ct}}
where $\mathbf{\bar H'}$ denotes the matrix with elements $H'_{m,n}$ and $\mathbf{\bar C}(0)$ 
contains the initial correlations. The entanglement entropy is then given by
\eq{
S(t) = 
\sum_l \ln(1 + \ee^{-2\omega_l(t)}) +
\sum_l \frac {2\omega_l(t)}{\ee^{2\omega_l(t)} + 1} =
\sum_l H(\zeta'_l(t))
\label{eq:ent}}
where $H(x)=-x\ln x -(1-x)\ln(1-x)$. The first (second) form proves to be useful 
for quenches from equal (unequal) fillings.

\section{Quench from equal fillings}

In this case, the initial correlation matrix has the block form 
\eq{
\mathbf{\bar C}(0) =
\left(
\begin{array}{cc}
{\bf C}^0 & 0 \\
0 & {\bf C}^0
\end{array}
\right)
}
where the $L \times L$ matrix $\mathbf{C}^0$ refers to a half-chain in the ground 
state and has elements
\eq{
C^0_{mn} = \sum_{k=1}^{L} \phi^0_k(m) \phi^0_k(n) \, n_k
\label{eq:cnull}
}
Here $n_k$ is the occupation number of mode $k$ and the wavefunctions
$\phi^0_k(m)$ have the form (\ref{eq:eigv}) with $2L$ replaced by $L$.
To evaluate (\ref{eq:ct}) one expresses the exponential operators as
\eq{
\left(\ee^{\pm i \mathbf{\bar H}'t}\right)_{mn}=\sum_{k=1}^{2L} \phi'_k(m) \phi'_k(n) \ee^{\pm i\Omega_k t}
\label{eq:expop}}
Using (\ref{eq:eigvcd}), defining the overlap matrix
\eq{
B_{kl} = \sum_{j=1}^L \phi_k(j) \phi^0_l(j) = (-1)^{(k+l)} \sum_{j=L+1}^{2L} \phi_k(j) \phi^0_l(j-L)
\label{eq:olmb}}
and noting that $\alpha_k \alpha_{k'} + (-1)^{k+k'}\beta_k \beta_{k'} = 2$ for $k-k'$ even 
and zero otherwise, one arrives at  
\eq{
C'_{mn}(t) = 2\sum_{\substack{k,k'=1 \\ k-k' \mathrm{even}}}^{2L}\sum_{k''=1}^{L}
%C'_{mn}(t) = \sum_{k,k''=1}^{2L}\sum_{k'=1}^{L}
\alpha_k \alpha_{k'}
% (\alpha_k \alpha_{k''} + (-1)^{k+k''}\beta_k \beta_{k''})
B_{kk''} B_{k'k''} \phi_k(m) \phi_{k'}(n) n_{k''} \ee^{i(\Omega_k-\Omega_{k'})t}
\label{eq:ctmn}}
The only dependence on the defect strength $\lambda$ is in the factors
$\alpha_k \alpha_{k'}$. The point now is that one can find a simple connection
with the homogeneous quench if one considers the matrix $(2\mathbf{C}'(t)-1)^2$.
Then a further overlap matrix appears with elements
\eq{
A_{kl} = \sum_{j=1}^L \phi_k(j) \phi_l(j)=
\frac{1}{2(2L+1)}\left [
\frac{\sin \frac{\pi}{2}(k-l)}{\sin \frac{\pi (k-l)}{2(2L+1)}} -
\frac{\sin \frac{\pi}{2}(k+l)}{\sin \frac{\pi (k+l)}{2(2L+1)}} \right]
\label{eq:olma}
}
which, apart from the diagonal ones, vanish for $k-l$ even. The matrix elements of
$(2\mathbf{C}'(t)-1)^2$ are 6-fold sums where each summand is proportional to
$\alpha_k \alpha_{k'} \alpha_l \alpha_{l'} A_{k'l}$ with
$k-k'$ and $l-l'$ even. For $k'-l$ odd, this gives
$\alpha_k \alpha_{k'} \alpha_l \alpha_{l'}=\lambda^2$ and thus
a simple prefactor. For $k'=l$, one has $A_{ll}=1/2$ and a summand
\eq{
%\sum_{\substack{k,k'=1\\ k-k'  \mathrm{even}}}^{2L}
%\sum_{\substack{l,l'=1 \\ l-l' \mathrm{ even}}}^{2L}
%\sum_{k'',l''=1}^{L}
%
\alpha_k \alpha^2_l \alpha_{l'}
B_{kk''}B_{lk''}B_{ll''}B_{l'l''}
\phi_k(m) \phi_{l'}(n) (2n_{k''}-1)(2n_{l''}-1)
\ee^{i(\Omega_k-\Omega_{l'})t}
}
where $\alpha_k \alpha^2_l \alpha_{l'} =2-\lambda^2 + (-1)^l2\sqrt{1-\lambda^2}$
for the allowed indices. The remaining sums can then be evaluated using the identities
\eq{
\sum_{l=1}^{2L} B_{lk''}B_{ll''} = \delta_{k''l''} , \quad
\sum_{l=1}^{2L} (-1)^l B_{lk''}B_{ll''} = 0 , \quad
\sum_{k''=1}^L B_{kk''} B_{l'k''}=A_{kl'}
}
Summing over $l$ therefore eliminates the alternating terms and enforces $k''=l''$
which leads to $(2n_{k''}-1)^2=1$ independent of $k''$ and of the filling.
The final sum over $k$ gives $(1-\lambda^2/2)\delta_{mn}$.
In this way one arrives at the formula
\eq{
(2\mathbf{C}'(t)-1)^2_{mn} = \lambda^2 (2\mathbf{C}(t)-1)^2_{mn}  +
(1-\lambda^2) \delta_{mn}
\label{eq:omeps}
}
As a consequence, the single-particle eigenvalues of the corresponding RDMs are
related by
$\tanh^2 \omega_l(t) = \lambda^2 \tanh^2 \varepsilon_l(t) + 1-\lambda^2$,
or alternatively, writing $\lambda=s$ for the transmission amplitude, by
\eq{
\ch \omega_l(t) =  \frac{1}{s} \ch \varepsilon_l(t) 
\label{eq:dispersion}
}
This is exactly the same relation one finds for the static defect problem \cite{Eisler/Peschel10}
and gives a gap in the $\omega_l$-spectrum. Note that (\ref{eq:omeps}) can also be written as
$\mathbf{C}'(t)(1-\mathbf{C}'(t)) = \lambda^2 \mathbf{C}(t)(1-\mathbf{C}(t))$
and is then identical to the relation for the overlap matrix $\mathbf{A}$ in a (static) continuum system
\cite{CMV11,CMV12a,CMV12b}. 

%%%%%%%%%%%%%%%%%%%%%%%%%%%%%%%%%%%%%%%%%%%%%%%%%%%%%%%%%%%
%
\begin{figure}[htb]
\center
\psfrag{(L)}[][][.6]{($\ell$)}
\psfrag{t , L}[][][.6]{t , $\ell$}
%\includegraphics[scale=0.6]{figs/eps_quench_vs_eq_l_n200.eps}
%\caption{Low-lying eigenvalues $\varepsilon_l(t)$ after homogeneous local quench
%and $\varepsilon_l(\ell)$ of an interval in a ring of size $2L=400$. The data is shown
%for half-filling.}
\includegraphics[scale=0.6]{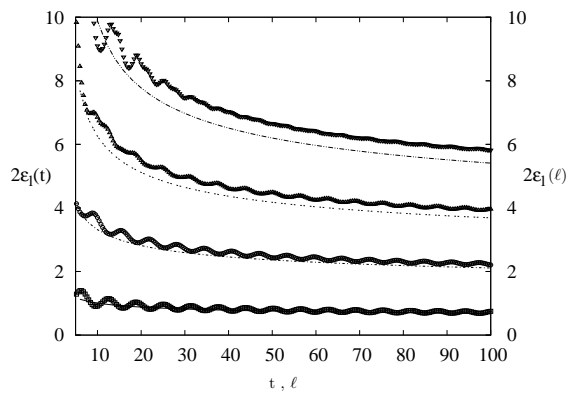}
\caption{Low-lying eigenvalues $2\varepsilon_l(t)$ after homogeneous local quench with
$2L=400$ (points) and $2\varepsilon_l(\ell)$ of an interval in an infinite system (lines).
The data is shown for even $\ell$ and for half-filling.}
\label{fig:eps}
\end{figure}
%
%%%%%%%%%%%%%%%%%%%%%%%%%%%%%%%%%%%%%%%%%%%%%%%%%%%%%%%%%%%

The problem is now reduced to that of the homogeneous quench, but one still needs the
$\varepsilon_l(t)$. The known results give a logarithmic behaviour of $S(t)$ with
a prefactor $1/3$ that is similar to the scaling of the equilibrium entropy of an interval $\ell$
in an infinite chain \cite{CC09}. In the static case, one knows that the low-lying $\varepsilon_l(\ell)$
have a spacing $\pi^2/2\ln \ell$ for large $\ln \ell$ \cite{Peschel04}. This gives a density of states
proportional to $\ln \ell$ and the logarithmic variation of $S(\ell)$. The similarity
of the conformal results for the entropy suggests that the $\varepsilon_l(t)$ have an analogous behaviour. 
This is in fact the case and is shown in Fig. \ref{fig:eps}.
Plotted are the lowest $\varepsilon_l(t)$ in a homogeneous quench together with the 
lowest $\varepsilon_l(\ell)$ for a segment of length $\ell$ in an infinite chain. Apart from 
some oscillations, they coincide closely. 

\begin{figure}[htb]
\center
\psfrag{T , s}[][][.65]{$T \, , \, s$}
\psfrag{c(s)}[][][.65]{$\hat c_{\mathrm{eff}}(s)$}
\psfrag{c(T)}[][][.65]{$\hat c_{\mathrm{eff}}(T)$}
\includegraphics[scale=0.6]{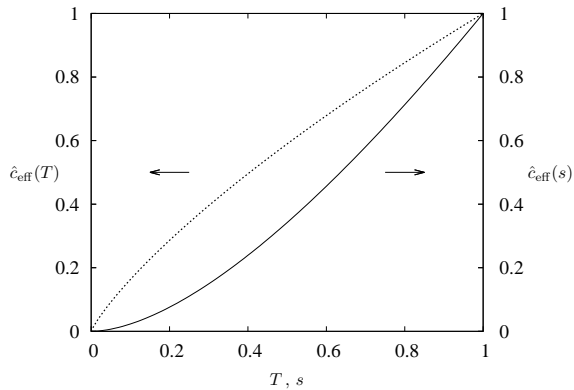}
\caption{$\hat c_{\mathrm{eff}}$ as a function of $s$ (solid line) as well as $T=s^2$ (dashed line).}
\label{fig:ceff}
\end{figure}

With this observation, one can now follow the same steps as in the static case and write
$S$ in (\ref{eq:ent}) as an integral over $\varepsilon$. In this way one finds 
$\hat c_{\mathrm{eff}}= c_{\mathrm{eff}}$, where $c_{\mathrm{eff}}$ is $12/\pi^2$ times the 
integral $I(s)$ given in (26) of \cite{Eisler/Peschel10}. It is depicted on Fig. \ref{fig:ceff},
both as a function of $s$ and $T=s^2$, to allow simple comparison with the numerical
results of \cite{Klich/Levitov09}. The function is nonanalytic around $s=0$.
For the simple bond defect, 
one knows from the static case that the parameter $s$ is the transmission amplitude at 
the Fermi level. The relation between $\omega$ and $\varepsilon$ is then only valid
as the levels become dense.

%%%%%%%%%%%%%%%%%%%%%%%%%%%%%%%%%%%%%%%%%%%%%%%%%%%%%%%%%%%
%
%
%%%%%%%%%%%%%%%%%%%%%%%%%%%%%%%%%%%%%%%%%%%%%%%%%%%%%%%%%%%
 
\section{Quench from unequal fillings}

If the initial fillings are unequal, a steady particle-current results after connecting
the half-chains. In the homogeneous case, the time evolution of
the density profile \cite{ARRS99}, the particle number fluctuations in a half-chain
\cite{AKR08} as well as the entanglement entropy \cite{EIP09} have been studied previously.
The fluctuations and the entropy both grow logarithmically in time.

\begin{figure}[htb]
\center
\includegraphics[scale=0.6]{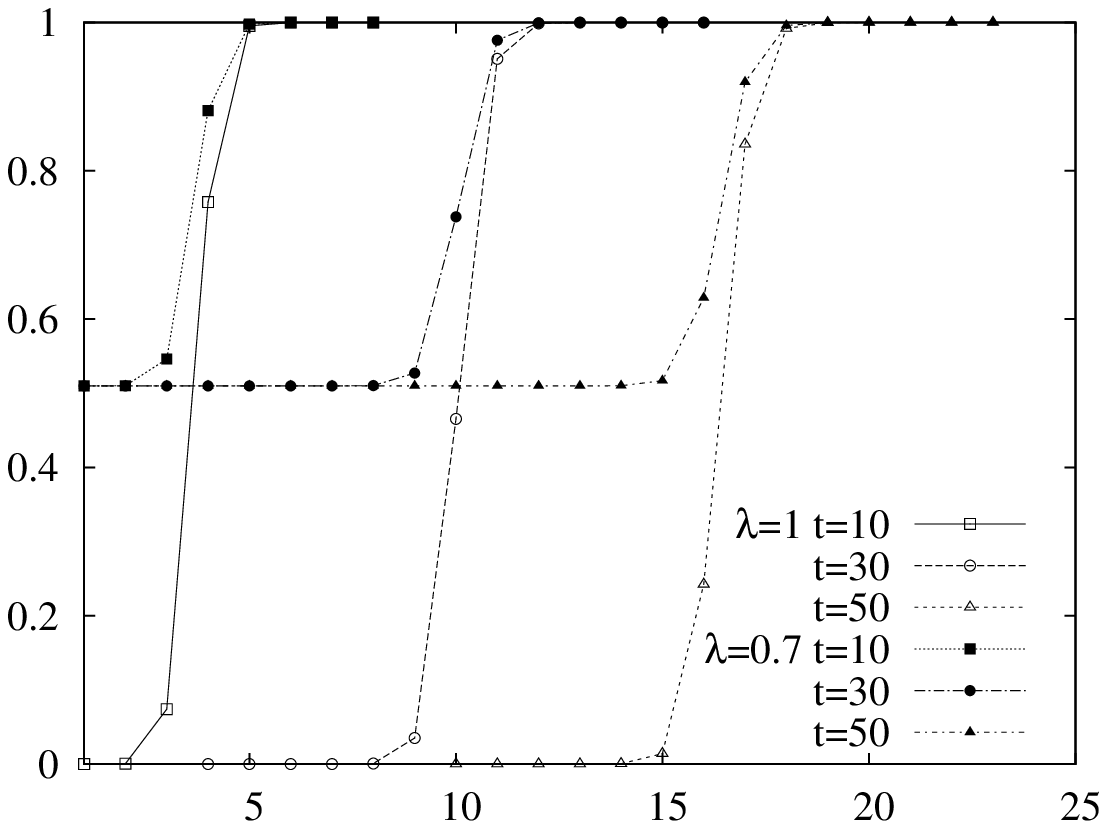}
\includegraphics[scale=0.6]{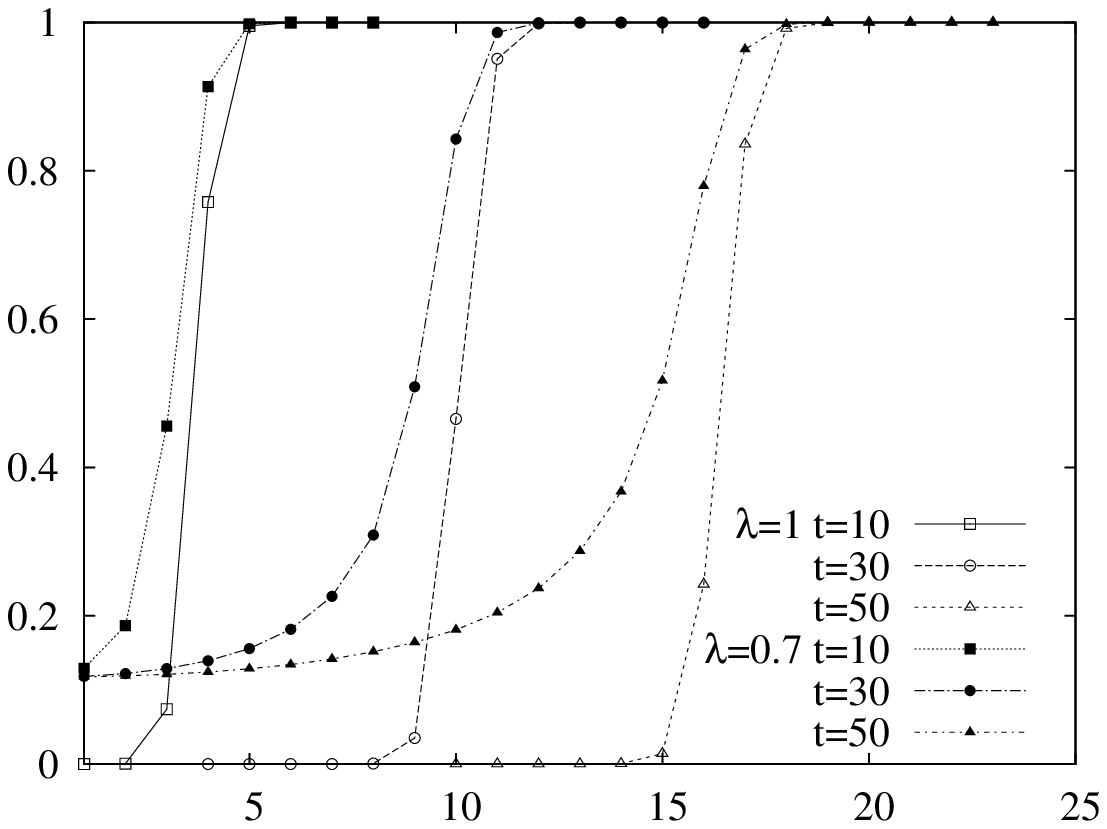}
\caption{Eigenvalues $\zeta_l(t)$ (open symbols) and $\zeta'_l(t)$ (filled symbols) for 
different times and $2L=200$ sites.
Left: conformal defect. Right: bond defect.}
\label{fig:zeta}
\end{figure}

We first consider the case where the left hand side of the chain
is completely filled and the right hand side is empty. The initial correlation matrix then reads
\eq{
\mathbf{\bar C}(0) =
\left(
\begin{array}{cc}
{\bf 1} & 0 \\
0 & 0
\end{array}
\right)
}
where $\mathbf{1}$ denotes the $L \times L$ identity matrix.
Expanding into eigenvectors one has
\eq{
C'_{mn}(t) = \sum_{k,l=1}^{2L}
\alpha_k^2 \alpha_l^2
\phi_k(m) A_{kl} \phi_l(n) \ee^{i(\Omega_k-\Omega_l)t} =
\lambda^2 C_{mn}(t) + (1-\lambda^2) \delta_{mn}
}
where in the second step we used the property $\alpha_k^2 \alpha_l^2 =\lambda^2$
for all $k-l$ odd, where the overlap matrix $\mathbf{A}$ is nonvanishing.
For $k=l$, the alternating piece in $\alpha_k^4$ gives zero upon summation and 
the constant piece leads to the second term on the right hand side.
Remarkably, one finds the same relation as in Eq. (\ref{eq:omeps})
but now for the correlation matrix itself and thus also for its eigenvalues
\eq{
\zeta'_l(t) = \lambda^2 \zeta_l(t) + 1-\lambda^2
\label{eq:zetarel}
}
The eigenvalues $\zeta'_l(t)$ and $\zeta_l(t)$ are shown on the left of Fig. \ref{fig:zeta}
and have a simple interpretation. As the half-chains are connected, particles leave the left 
part of the system and a steady current results. In the pure case, this implies a growing number of
$\zeta_l(t) = 0$ eigenvalues. In the presence of a defect, however, there is a probability 
$R=1-T$ for backscattering and the corresponding eigenvalues are $\zeta'_l(t) =1-\lambda^2=R$. 
It is interesting to compare with the case of a simple bond defect, which is done on the 
right of Fig. \ref{fig:zeta}. Here such a simple relation as (\ref{eq:zetarel}) does not hold, 
since the transmission coefficient depends on the momentum $q$ of the particle as
\eq{
T_q = \frac{\sin^2 q}{\ch^2 \nu -\cos^2 q}
\label{eq:trcxx}
}
where we defined $\lambda = \ee^{-\nu}$ and $0 \le q \le \pi$. Because of this variation,
the $\zeta'_l(t)$ curves are not flat.

%
%%%%%%%%%%%%%%%%%%%%%%%%%%%%%%%%%%%%%%%%%%%%%%%%%%%%%%%%%%%
%
%
%%%%%%%%%%%%%%%%%%%%%%%%%%%%%%%%%%%%%%%%%%%%%%%%%%%%%%%%%%%

\begin{figure}[htb]
\center
\includegraphics[scale=0.6]{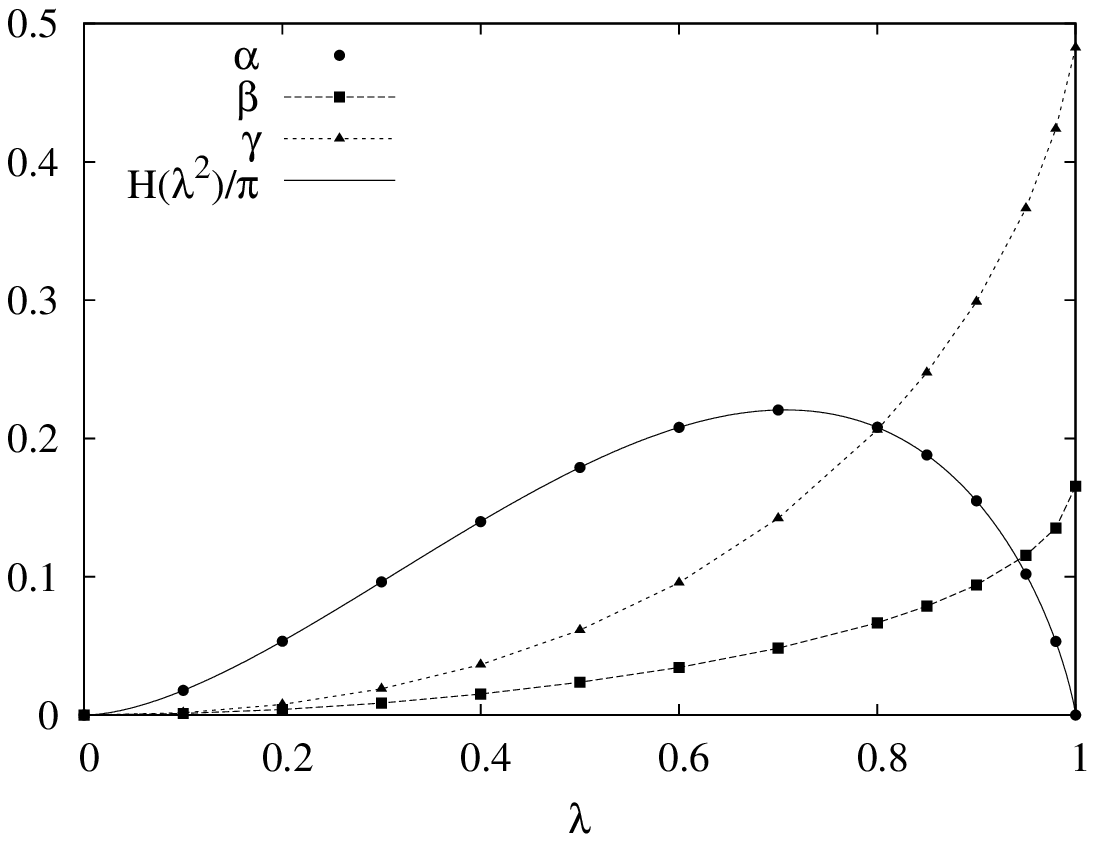}
\includegraphics[scale=0.6]{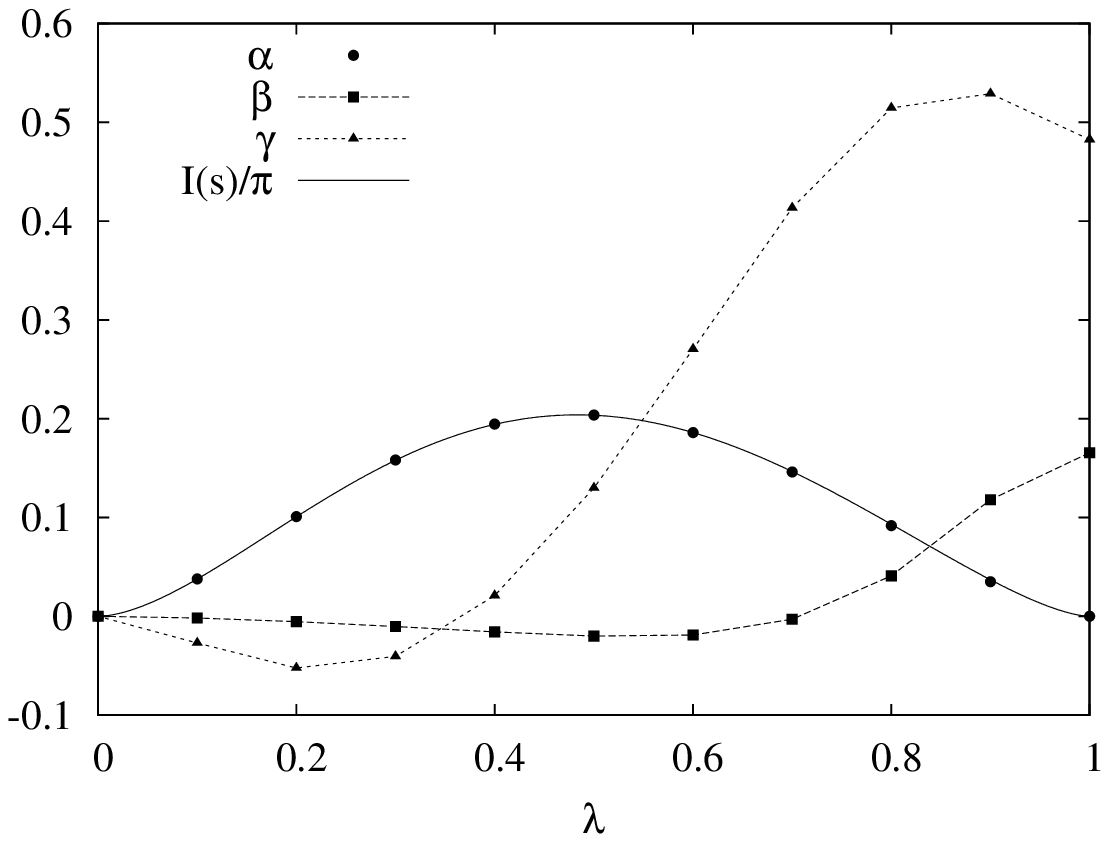}
\caption{Fitted coefficients of the ansatz $S(t) = \alpha t+ \beta \ln t+\gamma$
together with the analytical formulas obtained for the slope $\alpha$.
Left: conformal defect. Right: bond defect.}
\label{fig:coeff}
\end{figure}

The common feature of both defects is that the number of nonzero $\zeta'_l(t)$ eigenvalues,
which have a finite contribution to the entropy, grows with time proportionally to the number
of transmitted particles. Because of the steady flux in the center of the chain,
this results in a linear contribution to the entanglement.
Numerically, this can be observed even for coupling strengths arbitrarily close
to $\lambda=1$, where the entropy becomes logarithmic, $S \sim 1/6 \log t$ \cite{EIP09}.
In general, $S$ can be well fitted with the ansatz 
\eq{
S(t)=\alpha t+\beta \ln t + \gamma
\label{eq:fit}
}
and the coefficients are shown in Fig. \ref{fig:coeff}. In the conformal case, the linear 
part can be calculated, using the second form of the entropy in Eq. (\ref{eq:ent}), as 
$H(\lambda^2) \, t/\pi$. This is just the contribution of a single $\zeta'_l(t)$ on the 
flat part of the spectrum multiplied by the length of the plateau, which equals the total 
number $t/\pi$ of transmitted particles in the homogeneous case \cite{ARRS99}. 
This result, shown by the solid line on 
the left of the Fig. \ref{fig:coeff}, agrees perfectly with the fitted values of $\alpha$. 
Furthermore, it also agrees with the results of \cite{Klich/Levitov09,Song/etal11,Song/etal12}
for the entropy evolution in a quantum point contact in the high-bias regime.

%%%%%%%%%%%%%%%%%%%%%%%%%%%%%%%%%%%%%%%%%%%%%%%%%%%%%%%%%%%
%
%
%%%%%%%%%%%%%%%%%%%%%%%%%%%%%%%%%%%%%%%%%%%%%%%%%%%%%%%%%%%
%
%
%
In the non-conformal case, the linear growth of the entropy can be obtained by distinguishing
the different momentum states and associating a factor $H(T_q)$ with the entanglement which is
produced between transmitted and reflected parts of the wavefunction. Thus one writes for
$L \to \infty$
\eq{
\alpha=\int_0^{\pi} \frac{\dd q}{2\pi} v_q \, H(T_q)
\label{eq:aint}
}
where $v_q = d\Omega_q/dq = \sin q$ is the velocity of the incoming particle and measures the
flux.
For the conformal defect where $T=\lambda^2$ one obtains the previous result.
For the bond defect, the integral can be carried out by substituting
Eq. (\ref{eq:trcxx}) and using $T_q$ as integration variable. A lengthy calculation then 
yields $\alpha=I(s)/\pi$ with
\eq{
I(s)=\ln \left( \frac{1-s^2}{4s^2}\right) - \frac{1}{s}\ln \left( \frac{1-s}{1+s} \right)
+ \frac{1-s^2}{s} \left[\frac{1}{4}\ln^2 \left( \frac{1-s}{1+s} \right) 
+ \mathrm{Li}_2 \left( \frac{1-s}{1+s} \right) -\frac{\pi^2}{6} \right]
\label{eq:is}}
where $\mathrm{Li}_2 (s)$ denotes the dilogarithm function and
$s=1/\ch \nu=2/(\lambda+1/\lambda)$ is the same amplitude that appears in the equilibrium 
formulae. This analytical result is again in perfect agreement with the results of the data fits,
as shown on the right of Fig. \ref{fig:coeff}.

Note, that the structure of Eq. (\ref{eq:aint}) is very similar to the one found by
Fagotti and Calabrese for a \emph{global} quench in the XY model \cite{Fagotti/Calabrese08}.
However, in the present case entanglement is created only \emph{locally} (but steadily)
at the defect, in contrast with the global quench where entangled quasi-particle pairs are created
everywhere but only at $t=0$ \cite{CC05}. It would be interesting to check if this semiclassical
picture can be used to calculate other relevant observables as was found recently for
the global quench \cite{Rieger/Igloi11,BRI12}.

\begin{figure}[htb]
\center
\includegraphics[scale=0.6]{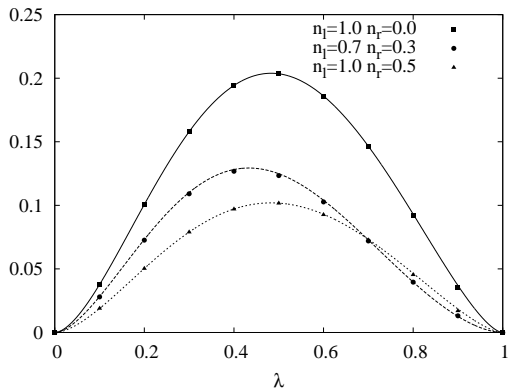}
\caption{Slope $\alpha$ of the linear part of the entropy for several filling factors as obtained by
fits to the data (points) as well as evaluating the integrals (lines)}
\label{fig:coeff2}
\end{figure}

The above arguments can be generalized to arbitrary initial filling factors $n_l$ and $n_r$ 
on the left and right. 
Assuming $n_l>n_r$, the single-particle states with $q < n_r \pi$ will be filled and thus
balanced on both sides. In terms of the $\zeta'_l(t)$, the number of eigenvalues on the flat
part of the spectrum decreases and one has a more general relation which interpolates
between (\ref{eq:zetarel}) and (\ref{eq:dispersion}).
In the semiclassical picture only wavenumbers $n_r \pi<q<n_l \pi$ will contribute to the
current and the integral in Eq. (\ref{eq:aint}) has to be carried out only on this interval.
As shown in Fig. \ref{fig:coeff2}, this gives again very good agreement with the numerical data,
further supporting the semiclassical picture described above.
It is possible to evaluate the integral analytically also in this case.
%%%%%%%%%%%%%%%%%%%%%%%%%%%%%%%%%%%%%%%%%%%%%%%%%%%%%%%%%%%
%
%
%%%%%%%%%%%%%%%%%%%%%%%%%%%%%%%%%%%%%%%%%%%%%%%%%%%%%%%%%%%
%

\section{Concluding remarks}

There is a close connection of our findings with the work on counting statistics 
by Klich and Levitov \cite{Klich/Levitov09}. For the quench with equal fillings,
they obtain a density of states for the eigenvalues $\zeta'$ which vanishes  
in an interval around $\zeta'=1/2$. Thus the spectrum has a gap, and this is just the 
situation described by (\ref{eq:dispersion}). Moreover, if one converts their entropy 
expression (10) into an integral over $\varepsilon$ one finds exactly the integral for 
$c_{\mathrm{eff}}$ evaluated in \cite{Eisler/Peschel10}. Finally, the relation between
their parameters $\lambda$ and $\lambda*$ is another form of the dispersion relation
(\ref{eq:dispersion}) derived here. Our approach, focussing on the eigenvalues themselves,
is more direct and shows the connection with the static defect problem very clearly. The only
open point is an analytical derivation of the $\varepsilon_l(t)$-behaviour, which we inferred
from conformal results and numerics.

In this context one should mention that we discussed only the case $t \ll L$.
The general CFT formula for the homogeneous quench is \cite{Stephan/Dubail11}
\eq{
S(t) = \frac{c}{3} \ln \left|\frac{2L}{\pi}\sin\frac{\pi v_F t}{2L}\right| + \mathrm{const.}
\label{eq:entconf}}
where the Fermi-velocity $v_F$ is set by the filling of the half-chains. This results again
from the scaling of the $\varepsilon_l(t)$ with the logarithmic factor and the formula for the 
defect case is obtained by substituting $c \rightarrow c_{\mathrm{eff}}$.

The relation $\hat c_{\mathrm{eff}} =  c_{\mathrm{eff}}$ also holds for the corresponding
coefficients in the R\'enyi entropies $S_n$ which are rather simple for $S_2$ and $S_3$
\cite{CMV12a,Peschel/Eisler12}. Similarly, the considerations for the biased quench
can be adapted to the R\'enyi functions. 

Finally, a quench from unequal fillings also arises in continuum models if one removes a wall
which initially confines the particles. A detailed study of the entanglement evolution
has appeared recently \cite{Vicari12} and can probably be extended to the defect case. 

\begin{acknowledgements}
We thank Pasquale Calabrese for an interesting discussion.
V.E. acknowledges financial support by the ERC grant QUERG.
\end{acknowledgements}

\section*{References}

\providecommand{\newblock}{}

\end{document}